\documentclass[showpacs,notitlepage,prd,aps,longbibliography,onecolumn]{revtex4-1}
\usepackage{graphicx}\usepackage{amsmath}\usepackage{amssymb}\usepackage{slashed}
\usepackage{dsfont}

\newcommand{\be}{\begin{equation}}\newcommand{\ee}{\end{equation}}
\newcommand{\bea}{\begin{eqnarray}}\newcommand{\eea}{\end{eqnarray}}
\newcommand{\nn}{\nonumber}
\newcommand{\pa}{\partial}
\newcommand{\ben}{\begin{enumerate}}\newcommand{\een}{\end{enumerate}}

\newcommand{\al}{\alpha}

\newcommand{\ep}{\varepsilon}
\newcommand{\om}{\omega}

\renewcommand{\kappa}{\varkappa}
\newcommand{\q}{\mathbf{q}}
\newcommand{\x}{\mathbf{x}}\newcommand{\y}{\mathbf{y}}\newcommand{\z}{\mathbf{z}}
\renewcommand{\a}{\mathbf{a}}
\newcommand{\p}{\mathbf{p}}\renewcommand{\k}{\mathbf{k}}

\newcommand{\n}{\mathbf{n}}\newcommand{\m}{\mathbf{m}}

\newcommand{\E}{\mathbf{E}}

\newcommand{\Ref}[1]{(\ref{#1})}

\newcommand{\elm}{electromagnetic~}
\newcommand{\od}{one dimensional~}
\newcommand{\td}{two dimensional~}
\newcommand{\dd}{three dimensional~}

\begin{document}
\title{Dirac lattices, zero-range potentials and self adjoint extension}
\author{M. Bordag\footnote{bordag@itp.uni-leipzig.de}, and J. M. Mu$\tilde{\rm n}$oz-Casta$\tilde{\rm n}$eda\footnote{jose.munoz-castaneda@uni-leipzig.de}\\
\footnotesize{{\sl Institut f\"ur Theoretische Physik, Universit\"at Leipzig, Germany.}}}
\date{\small \today,
}
\begin{abstract}
We consider the \elm field in the presence of polarizable point dipoles. In the corresponding effective Maxwell equation these dipoles are described by \dd delta function potentials. We review the approaches handling these: the selfadjoint extension, regularization/renormalisation and the zero range potential methods. Their close interrelations are discussed in detail and compared with the electrostatic approach which drops the contributions from the self fields.

For a homogeneous \td lattice of dipoles we write down the complete solutions, which allow, for example, for an easy numerical treatment of the scattering of the \elm field on the lattice or for investigating plasmons. Using these formulas, we consider the limiting case of vanishing lattice spacing, i.e., the transition to a continuous sheet. For a scalar field and for the TE polarization of the \elm field this transition is smooth and results in the results known from the continuous sheet. Especially for the TE polarization, we reproduce the results known from the hydrodynamic model describing a \td electron gas. For the TM polarization, for polarizability parallel and perpendicular to the lattice, in both cases, the transition is singular. For the parallel polarizability this is surprising and different from the hydrodynamic model. For perpendicular polarizability this is what was known in literature. We also investigate the case when the transition is done with dipoles described by smeared delta function, i.e., keeping a regularization. Here, for TM polarization for parallel polarizability, when subsequently doing the limit of vanishing lattice spacing, we reproduce the result known from the hydrodynamic model. In case of perpendicular polarizability we need an additional renormalization to reproduce the result obtained previously by stepping back from the dipole approximation.
\end{abstract}
%
%
\pacs{03.50.De}
\maketitle
%
\section{Introduction}\label{s1}
Recently, in \cite{para12-89-085021,bart13-15-063028,bord14-89-125015} the interaction of the \elm field with a monoatomically this sheet polarizable perpendicularly was considered and different results were found. As an attempt to understand theses results we start from a two dimensional lattice of polarizable point dipoles in \dd space and reconsider the standard treatment. On this way, for a scalar field one arrives at the equation
\be
    \left(-\om^2-\Delta+g\sum_\n \delta^3(\x-a_\n)\right)\Phi(\x)=0,
\label{1.1}\ee
where the points $a_\n$ ($\n=(n_1,n_2,0)$, $n_i$ integer) form a plane \td lattice of delta functions ('Dirac lattice').

Equations of such type, with delta function potential are known in different areas:
\ben
\item   The one-dimensional analog of eq. \Ref{1.1} with an equally spaced lattice, $a_n=a n$, is the well known Kronig-Penney model ('Dirac comb') \cite{kron31-130-499}, which now serves as a popular textbook example.
\item   In two and three dimensions, the delta functions in \Ref{1.1} are ill defined. For a single delta function this was observed in \cite{bere61-2-372} and solved by the method of self adjoint extensions.
\item   In quantum mechanics, putting $\om^2=2mE/\hbar^2$, eq. \Ref{1.1} is a Schr\"odinger equation with   delta function potentials. It appeared as generalization of the Kronig-Penney model and provides an approximation for scattering of short wave lengths particles. It is known as {\it method of zero-range potential} (see \cite{DemkovOstrovskii} and references therein). Also, sometimes it is called {\it point interaction}.
\item   In quantum mechanics, in two or three dimensions, using methods known from quantum field theory, this equation is considered with some regularization supplemented by subsequent renormalization of the coupling $g$. This was discussed in \cite{jack91} and, more recently in \cite{altu06-47-082110}, in application to delta functions on a manifold using a regularization in terms of separable potentials and application of renormaliation group techniques. A finite dimensional analog of this is known as Koster-Slater perturbation technique \cite{kost54-94-1167}.
\item   In classical electrodynamics with appropriate expression for $g$ (see below), this equation describes the electric field in the presence of   point dipoles, polarized by an applied field and by the field from all other dipoles. In this case, renormalization is equivalent to removing the self force.
\een
It must be mentioned that in each area the above mentioned methods are well developed, however, the communities using these, are quite disjunct. We find it useful to represent these methods together on one place demonstrating their close interrelation. We keep formulas as simple as possible to facilitate broader understanding and restrict ourselves to the examples relevant to the discussion of the polarizable sheets.

The recent discussion of thin polarizable sheets started with \cite{para12-89-085021,milt13-36-193}. For a polarizability perpendicular to the sheet no interaction with the electromagnetic field was found. In \cite{bart13-15-063028} a response was found starting from a plane lattice of dipoles polarizable perpendicularly to the sheet. However, for a shrinking of the lattice spacing to zero, it became singular. In \cite{bord14-89-125015}, also starting from a lattice, the limit of vanishing lattice spacing was taken in the equations. For polarizability parallel to the sheet, equations result which are equivalent to those of the hydrodynamic model, studied   in \cite{BV}. These are of the type of eq.\Ref{1.1}, however with a \od delta function, $\delta(z)$, and have a well defined, unique solution. For perpendicular polarizability, however, an equation results, which has in addition a term with the second derivative of the delta function, $\delta''(z)$. In order to match this term in the equation, the ansatz for the solution must itself contain a delta function. This is physically sound since the field $\Phi(\x)$ describes in this case the normal component of the electric field of a double layer which is know to have a delta function on the sheet. However, in this case the product $\Phi(\x)\delta(z)$ is singular. As a way out, in \cite{bord14-89-125015} it was discussed that one needs to step back from the underlying dipole approximation. This way, a finite result was obtained, which is different from \cite{para12-89-085021} and from \cite{bart13-15-063028}.
It is an aim of the present paper, to discuss the limit of vanishing lattice spacing in a broader context and to gain more insight into the different limiting procedures involved.

In electrodynamics, we consider the electric field in the presence of polarized point dipoles. The corresponding formulas are well known (see, e.g., \cite{Jackson75}). Here we follow the notations used in \cite{bord14-89-125015}. Putting $c=1$ and assuming harmonic time dependence $\sim\exp(-i\om t)$, we have 2 equations,
\bea    (-\om^2-\Delta)\E(\x)&=&4\pi(\om^2+\nabla\circ\nabla)\sum_\n \p_\n \delta(\x-\a_\n),
\nn\\    \p_\n&=&\hat{\al}\E(\a_\n),
\label{1.2}\eea
where the first line is the equation for the electric field in the presence of the point dipoles $\p_\n$ located at $\a_\n$, and the second line describes the polarization of one dipole by the electric field at the location of the dipole. In this way, the commonly used dipole approximation is assumed. In eq. \Ref{1.2}, $\nabla\circ\nabla$ is the dyadic product of the gradients. In general, all vectors are denoted by bold letters, for instance
\be \x=(x,y,z)=(\x_{||},z),\qquad \x_{||}=(x,y),
\label{1.3}\ee
and similar for the momenta. The sheet is always in the (x,y)-plane and the lattice
\be \a_\n=a\n,\qquad \n=(n_1,n_2,0)
\label{1.4}\ee
is quadratic with spacing $a$. In the second line in \Ref{1.2}, $\hat{\al}$ is the polarizability. It has the dimension of $length^3$. We take it as a diagonal matrix and consider two cases,
\bea
       \hat{\al}&=&{\rm diag}(\al_{||},\al_{||},0) ~~~~  \mbox{for in-plane polarizability},
\nn\\[4pt]   \hat{\al}&=&{\rm diag}(0,0,\al_3) ~~~~~~ \mbox{for perpendicular polarizability}.
 \label{1.5}
\eea
If thinking of the dipoles as bound charges $e$ with non relativistic motion, for $\al_{||}$ and $\al_3$ known expressions of the type
\be \al=\frac{e^2}{m(\om_0^2-\om^2)}
\label{1.6}\ee
with intrinsic frequency $\om_0$ hold in the simplest case. It is to be mentioned that the sign may change in dependence on the frequency. For $\om_0=0$ we get the polarizability for free charges as assumed in the hydrodynamic model \cite{BV}. However, in this paper we do not use these details.

Next we insert the dipole moments from the second line in \Ref{1.2} into the first line and we obtain an equation like \Ref{1.1},
\bea    \left(-\om^2-\Delta-4\pi(\om^2+\nabla\circ\nabla)\hat{\al}\sum_\n   \delta(\x-\a_\n)\right)\E(\x)=0.
\label{1.6a}\eea
and get an {\it effective Maxwell equation}.
It must be mentioned, that this equation is a direct consequence of the equations \Ref{1.2}. However, it is common not to use this equation in electrodynamics, see for example Chapt. 17 in \cite{Jackson75}. Below we will comment  more on this point.
Keeping aside for a moment the difficulties with the singularity, this is a vector equation. It can be diagonalized since the polarizations separate for the polarizabilities \Ref{1.5} and the geometry \Ref{1.4} into TE and TM modes,
\be \E(\x)=\left(\begin{array}{c}-\pa_y\\ \pa_x\\ 0\end{array}\right) E_{\rm TE}(\x)
            +\left(\begin{array}{c}\pa_x\pa_z\\ \pa_y\pa_z\\ -\Delta_{||}\end{array}\right) E_{\rm TM}(\x)
 \label{1.7}\ee
with $\Delta_{||}=\pa_x^2+\pa_y^2$. For in-plane polarizability we get an equation for the TE mode where $\nabla\circ\nabla$ does not contribute, whereas for the TM mode it becomes the in-plane Laplace operator, $\nabla\circ\nabla\to\Delta_{||}$. For perpendicular polarizability we have for parity reasons only the TM mode. In the following we denote this case be 'P'. We consider the equation for the normal component of the electric field, $\E_3(\x)$, and here we have to substitute $\nabla\circ\nabla\to\pa_z^2$. Together, in all three cases we get the modification
\bea    \left(-\om^2-\Delta+gP\sum_\n   \delta(\x-\a_\n)\right)\Phi(\x)=0.
\label{1.6b}\eea
of equation \Ref{1.1} or \Ref{1.6a} where for $g$ and $P$ one needs to substitute   according to 
\be
\begin{array}{c|c|rcl|c}
{\rm mode}&\multicolumn{4}{c|}{\rm coupling}&{\rm field}\\ \hline
TE&g\to\al_{||} ,& P &=&-4\pi \om^2,     &\Phi(\x)\to E_{\rm TE}(\x),\\
TM&g\to\al_{||} ,& P &=&-4\pi(\om^2+\Delta_{||}),&\Phi(\x)\to E_{\rm TM}(\x),\\
P&g\to\al_{3}   ,& P &=& -4\pi(\om^2+\pa_z^2),  &\Phi(\x)\to E_{3}(\x).
\end{array}\label{1.8}\ee
We mention that the TE mode differs from the scalar case only by the substitution of the coupling constant, while the other two cases have in addition derivatives in their couplings.

The paper is organized as follows. In the next section we consider the equation with a single delta function and discuss in detail the singularity, the regularization and  renormalization, which are used in section III for multiple centers. As an application we consider then the scattering on a two dimensional lattice. In section V we discuss the transition to a continuous sheet with different orders of the limits involved.

\section{Approaches for a single \dd  delta function}\label{s2}
In this section we consider the approaches to eq. \Ref{1.1} for a single delta function located at the origin, $\a_\n=0$,
\be
    \left(-\om^2-\Delta+g\delta^3(\x)\right)\Phi(\x)=0,
\label{2.1}\ee
where  $\delta(\x)=\delta(x)\delta(y)\delta(z)$ is a \dd delta function, $\x=(x,y,z)\in\mathbb{R}^3$ and $\Delta$ is the \dd Laplace operator.
As it stands, this equation is not well defined. In three (and in two) dimension this equation can be given a precise meaning by several the methods, which we review in this section.
\subsection{Self adjoint extension}\label{s2.1}
We begin with the method of self adjoint extensions. This was first done in \cite{bere61-2-372}, later generalized in \cite{Albeverio} and quite a number of further places. The idea is, roughly speaking, to restrict the domain of the operator to regular solutions where it is only symmetric, and then to add a singular solution which makes the operator self adjoint. In doing so, a new parameter $\al_{\rm SE}$ appears which can be called {\it extension parameter} and the original one, $g$, looses any meaning. In this way, any potential with support in one point is described. This solution appears only in the s-wave (orbital momentum $l=0$). Higher $l$ are not affected (these 'do not feel' the delta function) since any regular solutions behave $\sim r^l$ for $r\to0$. In this way, the solution has an expansion
\be
    \Phi(\x)=c\left(\frac{1}{|\x|}+\al_{\rm SE}+O(x)\right),
\label{2.1.1}\ee
where $c$ is an overall normalization constant and $\al_{\rm SE}$ is the extension parameter.
This extension is unique in the s-wave sector (see Sect. \ref{s2.2}).
Another,  equivalent form of writing,
\be
    \lim_{|\x|\to0} \left(-\al_{\rm SE}+\frac{{\rm d}}{{\rm d}|\x|}\right)|\x|\,\Phi(\x)=0,
    \quad\mbox{or}\quad \al_{\rm SE}=\lim_{|\x|\to0}\frac{{\rm d}}{{\rm d}|\x|}\ln\left(|\x|\Phi(\x)\right),
\label{2.1.2}\ee
is in terms of a boundary condition at the origin.

The solution to eq. \Ref{2.1} for $\x\ne0$, i.e., outside the delta function, can be written in the form
\be
    \Phi(\x)=e^{i\k\x}+f\,\frac{e^{i\om|\x|}}{|\x|}.
\label{2.1.3}\ee
It can be interpreted as incoming plane wave with wave vector $\k$ and an outgoing spherical wave with frequency $\om$ (with $\om=|\k|$ from eq. \Ref{2.1}), which is centered around the delta function. In this setup,  $f$ is the scattering amplitude. Expanding this solution,
\be\     \Phi(\x)=1+\frac{f}{|\x|}+i\om f+\dots\,,
\label{2.1.4}\ee
and comparing with eq. \Ref{2.1.1}, we identify
\be
\al_{\rm SE}=\frac{1}{f}+i\om,\quad\mbox{or},\quad f=\frac{1}{\al_{\rm SE}-i\om},
\label{2.1.5}\ee
which is a relation between the extension parameter $\al_{\rm SE}$ and the scattering amplitude. In this way, the solution \Ref{2.1.3} is extended to the whole plane, i.e., including $\x=0$.

The scattering amplitude $f$ has the appropriate analytic properties. It is a meromorphic function with a single pole on the  imaginary axis in $\om=i(-\al_{\rm SE})$, which for $\al_{\rm SE}<0$ corresponds to a bound state. The normalized boundstate wave function is
\be \phi_{\rm bs}(\x)=\sqrt{\frac{-\al_{\rm SE}}{2\pi}}\ \frac{e^{\al_{\rm SE} |\x|}}{|\x|}
\label{2.1.6}\ee
and $(-\al_{\rm SE})$ is the binding energy.

\subsection{The \dd delta function in the general theory of selfadjoint extensions}\label{s2.2}
In the last years Asorey, Munoz-Castaneda {\it et al} developed a new formalism to characterise the most general case of self adjoint extensions for operators relevant in mathematical physics in terms of an unitary operator $U$ (see Refs.
\cite{asor05-20-1001,asor13-874-852,asor08-41-304004,muno11-44-415401,muno14}).  
Here, we apply this formalism to a \dd delta function, which was not done so far.

In general, if $M$ is a $d$-dimensional manifold with boundary $\partial M\equiv\Sigma$, the set of selfadjoint extensions of the Laplace operator on $M$ is in one-to-one correspondence with the unitary group ${\cal U}\left(L^2(M)\right)$. This correspondence allows to characterise each self adjoint extension by an unitary operator $U$. In particular, for a given unitary operator $U\in{\cal U}\left(L^2(M)\right)$, the corresponding self adjoint extension $\Delta _U$ is fully determined by the domain of functions ${\cal D}(\Delta_U)\subset L^2(M)$ that satisfy the boundary condition
\be
\mu \left.\psi\right\vert_\Sigma-i\left.\partial_n\psi\right\vert_\Sigma=
U(\mu\left.\psi\right\vert_\Sigma+i\left.\partial_n\psi\right\vert_\Sigma),
\label{b.1}\ee
where the constant $\mu$ is a positive coupling constant with units of $L^{-1}$ that characterises the interaction with the boundary $\Sigma$. When the size of the boundary is finite, the coupling constant $\mu$ can be set to unity without loss of generality since the spectrum of $\Delta_U$ and its eigenfunctions only depend on the unitary operator $U$ that characterises the self adjoint extension (see Refs. \cite{asor13-874-852,asor08-41-304004}). In particular, the existence  of bound states in the spectrum of $\Delta_u$ only depends on the matrix $U$ and not in the value of $\mu$.

In the case of a singularity concentrated on a point as it is the delta function, one must introduce a regularisation in order to define the singular point-like potential. For this reason it can not be set to one. Assuming that the singularity is placed at the origin, the operator $U$  is defined on a sphere of radius $\epsilon$ around the singularity. Note that in such a case the boundary $\Sigma$ is a sphere of radius $\epsilon$ and the physical space of the system is given by the condition $r>\epsilon$. Therefore we have $\partial_n=-\partial_r$ and eq. \Ref{b.1} turns into
\be
\mu_\epsilon \left.\psi\right\vert_{r=\epsilon}+i\left.\partial_r\psi\right\vert_{r=\epsilon}=
U_\epsilon(\mu_\epsilon\left.\psi\right\vert_{r=\epsilon}-i\left.\partial_r\psi\right\vert_{r=\epsilon}).
\label{b.2}\ee
The radius $\epsilon$ is here the regularising parameter that must be made $0$ at the end. As it will be seen below, we need to allow the coupling constant $\mu$ to depend on the regularization parameter, $\mu\to\mu_\ep$. Therefore eq.\Ref{b.2} can be written as a boundary condition,
\be
\lim_{\epsilon\rightarrow0}\left(\mu_\epsilon \left.\psi\right\vert_{r=\epsilon}+i\left.\partial_r\psi\right\vert_{r=\epsilon}\right)=
\lim_{\epsilon\rightarrow0}U_\epsilon(\mu_\epsilon\left.\psi\right
\vert_{r=\epsilon}-i\left.\partial_r\psi\right\vert_{r=\epsilon}).
\label{b.3}\ee
When the singularity preserves spherical symmetry, the operator $U$, that characterises the corresponding self adjoint extension, is box-diagonal when the wave function is decomposed in spherical components. Therefore for each value of the orbital angular momentum $L$, we will have one boundary condition characterised by the a finite matrix $U^{(L)}_{M,M'}$ of order $2L\times 2L$ (see Refs. \cite{Munoz-Castaneda2009,ibort14}). Any quantum state $\psi_k({\bf x})$ with defined energy $k^2$ can be decomposed in spherical coordinates as
\be
\psi_k({\bf x})=\sum_{L=0}^\infty\sum_{M=-L}^L R_{k,LM}(r)Y_{LM}(\Omega),
\label{b.4}\ee
where $Y_{LM}(\Omega)$ are the spherical harmonics, and $R_{k,LM}$ is the radial function that obeys  the differential equation,
\be
-\frac{1}{r^2}\frac{d}{r}\left(r^2\frac{d}{r}R_{LM}(r)\right)+\frac{L(L+1)}{r^2}R_{LM}(r)=k^2R_{LM}(r).
\label{b.5}\ee
The general solution to the radial differential equation is given by
\be
R_{LM}(r)=A_{LM}(k)j_L(k r)+B_{LM}(k)y_L(kr),\label{b5-2}
\ee
where $j_L(r)$ and $y_L(r)$ are the spherical Bessel functions. Notice that since the boundary condition in general does not require regularity of the radial function at the origin, the coefficient of $y_L(k r)$ can not be made equal to zero. Hence, for fixed $L$, the corresponding boundary conditions have the form
\be
\left.\mu_\epsilon\psi\right\vert_{r=\epsilon}\pm i \left.\partial_r\psi\right\vert_{r=\epsilon}=\sum_{M=-L}^L(\mu_\epsilon R_{k,LM}(\epsilon)\pm i R^\prime_{k,LM}(\epsilon))Y_{LM}(\Omega)
\label{b.6}\ee
for any $L$. Since the spherical harmonics are orthonormal, these should be taken as the basis of a $2L$ vector space of boundary data for fixed $L$. The   boundary data can be represented as two $2L$-dimensional column vectors,
\be
\Phi^{(L)}_{\pm}\equiv\left(\begin{tabular}{c}
$\mu_\epsilon R_{k,L-L}(\epsilon)\pm i R^\prime_{k,L-L}(\epsilon)$ \\

$ \vdots$ \\

$\mu_\epsilon R_{k,LM}(\epsilon)\pm i R^\prime_{k,LM}(\epsilon)$\\

$ \vdots$ \\

$\mu_\epsilon R_{k,LL}(\epsilon)\pm i R^\prime_{k,LL}(\epsilon)$ \\
\end{tabular} \right).
\label{b.7}\ee
Now, the most general  boundary condition that preserves the total angular momentum can be written as
\be
\Phi^{(L)}_+={\bf U}_L\Phi^{(L)}_-\,,~~~~ \,\, {\bf U}_L\in U(2L+1),\,\,L=0,1,2,3,\dots \,.
\label{b.8}\ee
This  boundary condition is characterised by a set of matrices $\{ {\bf U}_L\vert\, {\bf U}_L\in U(2L+1)\}_{L\in\mathbb{Z}^+}$. Notice, that when the boundary condition is defined by ${\bf U}_L=\mathbb{I}_{2L+1\times2L+1}$, one obtains the boundary condition $ R^\prime_{k,L-L}(\epsilon)=0$ which is Neumann boundary condition. If ${\bf U}_L=-\mathbb{I}_{2L+1\times2L+1}$ gives rise to $\epsilon R_{k,LL}(\epsilon)=0$ which must be satisfied for any $\epsilon>0$ before taking the limit $\epsilon\rightarrow0$ which happens if and only if $R_{k,LL}(\epsilon)=0$ which is Dirichlet boundary condition. When taking the limit $\epsilon \rightarrow 0$ for the cases of Dirichlet and Neumann boundary conditions, the coefficient $B_{LM}$ in equation (\ref{b5-2}) must be equal to zero which gives rise to functions regular at the origin. It is of note that when the point potential is spherically symmetric, the matrices ${\bf U}_L$ must be diagonal and the boundary condition \Ref{b.8} becomes
\be
R^\prime_{k,LM}(\epsilon)=\mu_\epsilon\tan\left( \theta_\epsilon(L,M)/2\right)R_{k,LM}(\epsilon)
\label{b.9}\ee
for each radial function $R_{k,LM}$.

Within this formalism, it is also possible to define a multipole boundary condition by interpreting the $L$-decomposition of the  boundary condition given above as a multipole decomposition of the point potential. Therefore, when $U_L\neq-\mathbb{I}_{2L+1\times2L+1}$ just for a given $L_0\geq 0$, then one should interpret the point potential as the short range approximation for the potential of a $L$-multipole ($L_0=0$ would be the monopole or point charge, $L_0=1$ the dipole, $L=2$ the quadrupole, and so on). As it is well known the delta function potential  represents the short range approximation of a point charge (this is the same as the zero range approximation discussed in the next subsection). Therefore, to represent the delta function potential as a  boundary condition, only the monopole term in the multipole decomposition of the  boundary condition should be non-trivial and $U_L^{(\delta)}=-\mathbb{I}$ for all $L>0$.

To determine the $L=0$ component we have to take into account that in this case $U_{L=0}=e^{i \theta}\in U(1)$. Therefore the boundary condition for the $L=0$ components reads
\be
\mu_\epsilon R_{k,0}(\epsilon)+i R^\prime_{k,0}(\epsilon)=e^{i \theta_\epsilon}\mu_\epsilon R_{k,0}(\epsilon)-i R^\prime_{k,0}(\epsilon).
\label{b.10}\ee
To determine $\mu_\epsilon$ and $\theta_\epsilon$ in terms of $\epsilon$ and the extension parameter $\alpha_{SE}$ introduced in  \Ref{2.1.1}, we need to write down  condition \Ref{2.1.2} in a suitable manner. Previous to taking the limit $|{\bf x}|\rightarrow 0$ in equation (\ref{2.1.3}) we can write condition \Ref{2.1.2} over a small sphere of radius $\epsilon$,
\be
(1-\epsilon\alpha_{SE})R_{k,0}(\epsilon)+\epsilon R^\prime_{k,0}(\epsilon)=0\Rightarrow R^\prime_{k,0}(\epsilon)=\frac{\epsilon\alpha_{SE}-1}{\epsilon}R_{k,0}(\epsilon).
\label{b.11}\ee
Provided $e^{i\theta}\neq\pm 1$, we can write the condition \Ref{b.10} as
\be
R^\prime_{k,0}(\epsilon)=\mu_\epsilon\tan\left( \theta_\epsilon/2\right)R_{k,0}(\epsilon).
\label{b.12}\ee
Comparing the last two expressions we obtain the condition
\be
\mu_\epsilon\tan\left( \theta\epsilon/2\right)=\frac{\epsilon\alpha_{SE}-1}{\epsilon}\,.
\label{b.13}\ee
Keeping in mind that the only dimensional parameter entering the   boundary condition  \Ref{b.1} is $\mu_\epsilon$, we obtain the relations
\be
\mu_\epsilon=1/\epsilon,\qquad\tan\left( \theta_\epsilon/2\right)=\epsilon\alpha_{SE}-1,
\label{b.14}\ee
connecting the parameters of the self adjoint extension in subsection \ref{s2.2} and in the present subsection.

It is of note that the renormalization group equations for the parameters $\mu_\epsilon$ and $\theta_\epsilon$ are
\be
\frac{d \mu_\epsilon}{d \epsilon}=-1/\mu_\epsilon^2;\quad\frac{d \theta_\epsilon}{d \epsilon}=\frac{\alpha_{SE}}{1-\tan^2(\theta_\epsilon)}.
\label{b.15}\ee
This dynamical system also characterises the \dd delta function up to two integration constants.

The generalisation of the definition of the delta function potential for an electromagnetic field can be achieved using the theory of self-adjoint extensions for the quadratic Yang-Mills operator around a point-like configuration (see Refs. \cite{muno-notes,sandro-phd,ibort-lect}).

\subsection{Zero-range potential}\label{s2.3}
In quantum mechanics, eq. \Ref{2.1} with $\om=\sqrt{2mE}/\hbar$ is a Schr\"odinger equation for a particle with mass $m$ moving in a delta function potential. For a generic spherical symmetric potential $V(r)$ in place of the delta function one considers the scattering of the particle off the potential and comes to a partial wave scattering amplitudes $f_l(k)$ ($l=0,1,2,\dots$, $k=|\k|$). For small momenta $k$, or for large wave length of the scattered particle, these can be expanded,
\be f_l(k)=-a_l k^{2l}+\dots
\label{2.2.1}\ee
(see, e.g., \cite{tayl72b} chap. 11), where $a_l$ is the scattering length, which is a property of the potential. For small $k$, eq.\Ref{2.2.1} provides an approximation for the scattering amplitude. If restricting in addition to s-wave scattering, comparison with eq. \Ref{2.1.5} shows the relation
\be a_0=-\frac{1}{\al_{\rm SE}}
\label{2.2.2}\ee
between the extension parameter and the s-wave scattering length. For the s-wave, in the next term in expansion \Ref{2.2.1},
\be f_0(k)=-a_0+\frac12 r_0k^2+\dots,
\label{2.2.3}\ee
the parameter $r_0$ is interpreted as range of the potential. Putting $r_0=0$ motivates the naming 'method of zero-range potential'.

This method was actively used since the  mid 1960ies, see,  for example, \cite{DemkovOstrovskii} and also the more recent paper \cite{bond09-0907.2044}, which has an exhaustive bibliography.
Some most recent applications are \cite{borz13-88-033410} and \cite{coju14-47-315201}.
The equivalence to the method of self adjoint extension was shown in \cite{karp83-57-1231}.
%
%
\subsection{Regularization and renormalization}\label{s2.4}
In this subsection we consider another approach which follows a line of reasoning know in quantum field theory. Following \cite{jack91}, we start with a Fourier transform,
\be
    \Phi(\x)=\int\frac{d^3\p}{(2\pi)^3}\ e^{i\p\x}\ \tilde{\phi}(\p),
\label{2.3.1}\ee
and get from \Ref{2.1}
\be
    \left(-\om^2+\p^2\right)\tilde{\phi}(\p)+g\Phi(0)=0
\label{2.3.2}\ee
with $\om^2=\k^2$. This equation can be easily solved,
\be
    \tilde{\phi}(\p)=(2\pi)^3\delta(\p-\k)-\frac{g\Phi(0)}{p^2-\om^2},
\label{2.3.3}\ee
assuming $\Im \om>0$. The first term is the homogeneous solution. Inserting \Ref{2.3.3} into \Ref{2.3.1} we get for $\x=0$ the equation
\be
    \Phi(0)=1-g\Phi(0)I(-\om^2-i0),
\label{2.3.4}\ee
where
\be
    I(z)=\int\frac{d^3\p}{(2\pi)^3}\ \frac{1}{p^2+z}
\label{2.3.5}\ee
carries the divergence. This divergence is one way to see that the delta function in eq. \Ref{2.1.1} is ill defined in dimensions higher than one. At once, this integral gives the opportunity to introduce a regularization, i.e., to change the initial formal expression in a way that it becomes well defined. A convenient way is to change the dimension, $3\to n$, in the integration and, following the aim of dimensional regularization, to consider the analytic continuation of $I(z)$ in $n$. Another way is to restrict the integration, $|\p|<\Lambda$, by a momentum cut-off with $\Lambda\to\infty$ at the end. Thus one substitutes
\be
    I(z)\to I^{\Lambda}(z)
        = \int\limits_{|\p|<\Lambda}\frac{d^3\p}{(2\pi)^3}\ \frac{1}{p^2+z}
        = \frac{\Lambda}{2\pi^2}-\frac{\sqrt{z}}{4\pi}+O\left(\frac{1}{\Lambda}\right)
\label{2.3.6}\ee
and has explicitly the singularity for $\Lambda\to\infty$. Rewriting eq. \Ref{2.3.4} with this regularization and solving for $g\Phi(0)$,
\be
    g\Phi(0)=\frac{1}{\frac{1}{g}+\frac{\Lambda}{2\pi^2}+\frac{i\om}{4\pi}+\dots}\,,
\label{2.3.7}\ee
one defines by
\be
    \frac{1}{g_r}=\frac{1}{g}+\frac{\Lambda}{2\pi^2}
\label{2.3.8}\ee
a new, renormalized coupling constant $g_r$ which is supposed to be finite for $\Lambda\to\infty$. This is on expense of the initial coupling constant $g$, which is now $g=\left(\frac{1}{g_r}-\frac{\Lambda}{2\pi^2}\right)^{-1}$ and which goes to zero.
Finally, inserting $g\Phi(0)$ from \Ref{2.3.7} with \Ref{2.3.8} for $\Lambda\to\infty$ into \Ref{2.3.3} we get the solution which after Fourier transform back becomes
\be
 \Phi(\x)=e^{i\k\x}-\frac{1}{\frac{1}{g_{r}}+\frac{i\om}{4\pi}}\,\frac{e^{i\om|\x|}}{4\pi |\x|}.
\label{2.3.9}\ee
This is just the solution \Ref{2.1.5} with the identification
\be f=\frac{-1}{\frac{4\pi}{g_{r}}+i\om},\quad \mbox{resp.,}
\quad \al_{\rm SE}=\frac{4\pi}{g_r}\,.
\label{2.3.10}\ee
In this way, one obtains after renormalization the same result as from self adjoint extension, eq. \Ref{2.1.5}. The freedom in the renormalization (one can add to \Ref{2.3.8} any constant) corresponds to the freedom in the choice of the extension parameter $\al_{\rm SE}$.

At this place it should be repeated that the initial parameter $g$ in the process of renormalization looses its meaning completely. The new, renormalized parameter $g_r$ can be given a meaning by relating it via \Ref{2.3.10} with the scattering amplitude or, using \Ref{2.1.6} with a bound state level.

The cut-off regularization introduced with eq. \Ref{2.3.6} is equivalent to the use of a regularized delta function $\delta(\ep,\x)$ in eq. \Ref{2.1} which is non singular and has the property
\be \lim_{\ep\to0}\delta(\ep,\x)=\delta(\x).
\label{2.3.11}\ee
However, for any non zero $\ep$, one comes this way to an equation with a generic potential. Especially when considering several such potentials in an equation like eq. \Ref{1.1}, variables do not separate and the equation becomes hard to investigate. There is a way to avoid this difficulty by considering the regularized equation in the form
\be (-\om^2-\Delta)\Phi(x)+g\delta_1(\ep,\x)\int d\y \delta_2(\ep,\y) \Phi(\y)=0,
\label{2.3.12}\ee
where we introduced two regularized delta functions which may be different from one another.

Starting from here, we consider in parallel to the scalar case, eq. \Ref{1.1}, also the electric field obeying eqs. \Ref{1.2}, with s single dipole located at the origin, $\a_\n=0$.
We use this freedom in the choice of the regularized delta functions for accommodating the cases we are interested in.
Thus we demand
\be \lim_{\ep\to0}\delta_1(\ep,\x)=P\delta(\x),\qquad \lim_{\ep\to0}\delta_2(\ep,\x)=\delta(\x).
\label{2.3.12a}\ee
with $P=1$ for the scalar case and with $P$ either $P\to -4\pi(\om^2+\nabla\circ\nabla)$ (and $g\to -4\pi\hat{\al}$) or, according to formula \Ref{1.8}, for the electromagnetic case.
Clearly, for $\ep\to0$, using \Ref{2.3.12a}, we get back to eq. \Ref{2.1} for the scalar case. For the electric field we substitute $\Phi(\x)\to\E(\x)$ in \Ref{2.3.12} and get back eq. \Ref{1.6a}.

Rewriting eq. \Ref{2.3.12} in the form
\be \int d\y K(\x,\y) \Phi(\y)=0,
\label{2.3.13}\ee
with the integral kernel
\be K(\x,\y)=(-\om^2-\Delta)\delta(\x-\y)+g\delta_1(\ep,\x)\delta_2(\ep,\y),
\label{2.3.14}\ee
it is seen that we have an equation with a separable potential. Defining the Green function by
\be     \int d\z K(\x,\z)G(\z,\y)=\delta(\x-\y),
\label{2.3.15}\ee
the explicit solution is
\be G(\x,\y)=G_{0}(\x-\y)-\int d\x'd\y' G_{0}(\x-\x')\delta_1(\ep,\x')\phi_0^{-1}
                            \delta_2(\ep,\y')G_{0}(\y'-\y)
\label{2.3.16}\ee
with
\be \phi_0=\frac{1}{g}+\int d\x d\y \delta_2(\ep,\x)G_0(\x-\y)\delta_1(\ep,\y)
\label{2.2.17}\ee
and the free Green function obeying $(\om^2-\Delta)G_{0}(\x-\y)=\delta(\x-\y)$. Its explicit form is
\be G_{0}(\x-\y)=\int\frac{d\p}{(2\pi)^3}\ \frac{e^{i\p\x}}{\p^2-\om^2-i0}
                    =\frac{e^{i\om|\x|}}{4\pi|\x|}.
\label{2.3.17a}\ee
The solution \Ref{2.3.16} can be easily checked by inserting into \Ref{2.3.15}.

From \Ref{2.3.16}, a solution corresponding to an incoming plane wave with wave vector $\k$, can be obtained by
\be \Phi(\x)=\int d\y d\z \  G(\x,\y)G_{0}^{-1}(\y-\z)e^{i\k\z}
\label{2.3.18}\ee
and with \Ref{2.3.16} it can be written in the form
\be \Phi(\x)=e^{i\k\x}-
        \int d\x' G_{0}(\x-\x')\delta_1(\ep,\x')
        \phi_0^{-1}\int d\y'\delta_2(\ep,\y')e^{-i\k\y'}.
\label{2.3.19}\ee
For $\ep\to0$ it turns into the solution \Ref{2.1.3} with the identification
\be f=\lim_{\ep\to0}\frac{-1}{4\pi\phi_0}.
\label{2.3.20}\ee
This way, the equivalence of the separable regularization with the previously considered approaches is seen.

An especially convenient choice for the regularized delta function is the heat kernel,
\be 
        K_\ep(x)=\frac{\exp\left(-\frac{x^2}{4s}\right)}{(4\pi s)^{3/2}},
\label{2.3.21}\ee
with its property $\lim_{\ep\to0}K_\ep(\x)=\delta(\x)$ and the pleasant formulas
\bea G_0(\x)&=&\int_0^\infty ds \,e^{-s\xi^2}K_{s}(\x),
\nn\\\int d\z \, G_0(\x-\z)K_\ep(\z-\y) &=&\int_0^\infty ds \,e^{-s\xi^2}K_{s+\ep}(\x-\y),
\label{2.3.22}\eea
which follow from \Ref{2.3.17a} and \Ref{2.3.21}. The parameter $\ep$ regularizes the singularity in the integration over the proper time $s$ at $s=0$. We use imaginary frequency, $\om=i\xi$, in order to avoid inconvenience  with powers of $i$. We use the heat kernel regularization in the following way. For the scalar case we take $\delta_1(\ep,\x)=K_\ep(\x)$. For the polarizations of the electromagnetic case we take $\delta_1(\ep,\x)=PK_\ep(\x)$ which with \Ref{1.8} turns into
\be\begin{array}{lrcl}
{\rm TE:}~~~~~  & \delta_1(\ep,\x)  &=& 4\pi \xi^2 K_\ep(\x),   \\[4pt]
{\rm TM:}  & \delta_1(\ep,\x)  &=& 4\pi (\xi^2-\Delta_{||}) K_\ep(\x),\\[4pt]
{\rm P:}  & \delta_1(\ep,\x)  &=& 4\pi (\xi^2-\pa_z^2) K_\ep(\x),
\end{array}
\label{2.3.23}\ee
and for all cases we take $\delta_2(\ep,\x)=K_\ep(\x)$. Now we insert these definitions into eq. \Ref{2.2.17}. For the scalar case we get
\be \phi_0=\frac{1}{g}+\int_0^\infty ds\, e^{-s\xi^2}K_{s+2\ep}(0)
\label{2.3.24}\ee
and for the electromagnetic cases we get
\bea     \phi_0^{\rm TE} &=&
\frac{1}{\al_{||}}+4\pi\int_0^\infty ds \ e^{-s\xi^2}\xi^2K_{s+2\ep}(\x)_{\big| \x=0},
\nn\\        \phi_0^{\rm TM} &=&
\frac{1}{\al_{||}}+4\pi\int_0^\infty ds \ e^{-s\xi^2}(\xi^2-\Delta_{||})K_{s+2\ep}(\x)_{\big| \x=0},
\nn\\        \phi_0^{\rm P} &=&
\frac{1}{\al_{3}}+4\pi\int_0^\infty ds \ e^{-s\xi^2}(\xi^2-\pa_z^2)K_{s+2\ep}(\x)_{\big| \x=0}.
\label{2.3.25}\eea
Calculated for $\ep\to0$, these expressions become divergent according to
\bea      \phi_0^{\rm scalar} &=&
\frac{1}{g}+\frac{1}{(2\pi)^{3/2}\sqrt{\ep}}+ \frac{i\om}{4\pi}+O(\sqrt{\ep}),
\nn\\        \phi_0^{\rm TE} &=&
\frac{1}{\al_{||}}+-\frac{\om^2}{\sqrt{2\pi\ep}}+i\om^3+O(\sqrt{\ep}),
\nn\\        \phi_0^{\rm TM} &=&
\frac{1}{\al_{||}}+ +\frac{1}{6\sqrt{2\pi}\ep^{3/2}}-\frac{\om^2}{3\sqrt{2\pi\ep}}+\frac{i\om^3}{3}+O(\sqrt{\ep}),
\nn\\        \phi_0^{\rm P} &=&
\frac{1}{\al_{3}}+\frac{1}{12\sqrt{2\pi}\ep^{3/2}}-\frac{\om^2}{6\sqrt{2\pi\ep}}
                                +\frac{2i\om^3}{3}+O(\sqrt{\ep}),
\label{2.3.26}\eea
where we turned back to real frequencies, $\xi=-i\om$, and give rise to the renormalizations of the couplings,
\be\begin{array}{lrll}
    \frac{1}{g^{\rm ren}}   &=&  \frac{1}{g}+\frac{1}{(2\pi)^{3/2}\sqrt{\ep}}, &\quad {\rm for~ scalar,}\\[4pt]
    \frac{1}{\al_{||}^{\rm ren}}   &=&  \frac{1}{\al_{||}}-\frac{\om^2}{\sqrt{2\pi\ep}},&\quad {\rm for~ TE,}\\[4pt]
     \frac{1}{\al_{||}^{\rm ren}}   &=&  \frac{1}{\al_{||}}+\frac{1}{6\sqrt{2\pi}\ep^{3/2}}-\frac{\om^2}{3\sqrt{2\pi\ep}},&\quad {\rm for~ TM,}\\[4pt]
      \frac{1}{\al_{3}^{\rm ren}}   &=&  \frac{1}{\al_{3}}+\frac{1}{12\sqrt{2\pi}\ep^{3/2}}-\frac{\om^2}{6\sqrt{2\pi\ep}},&\quad {\rm for ~P.}
       \end{array}
\label{2.3.27}\ee
It is seen that all these renormalizations are different one from another and  depend in the electromagnetic case on frequency. In terms of the renormalized couplings, we get
\be \phi_0^{\rm scalar}=\frac{1}{g^{\rm ren}}+\frac{i\om}{4\pi},\quad
    \phi_0^{\rm TE}=\frac{1}{\al_{||}^{\rm ren}}+i\om^3,\quad
    \phi_0^{\rm TM}=\frac{1}{\al_{||}^{\rm ren}}+\frac{i\om^3}{3},\quad
    \phi_0^{\rm P}=\frac{1}{\al_{3}^{\rm ren}}+\frac{2i\om^3}{3}.
\label{2.3.27a}\ee
Finally we consider the wave functions which follow from eq. \Ref{2.3.19} with $\ep\to0$ and \Ref{2.3.17a}. In the scalar case we get
\be \Phi(\x)=e^{i\k\x}-\frac{1}{4\pi\phi_0^{\rm scalar}}\frac{e^{i\om|\x|}}{|\x|},
\label{2.3.28}\ee
repeating the identification \Ref{2.3.20} and eq. \Ref{2.3.28} coincides with eq. \Ref{2.1.3}.

In the electromagnetic cases we get, using \Ref{2.3.23},
\bea E_{\rm TE}(\x) &=&e^{i\k\x}+\frac{\om^2}{\phi_0^{\rm TE}}\frac{e^{i\om|\x|}}{|\x|},
\nn\\ E_{\rm TM}(\x) &=&e^{i\k\x}+\frac{1}{\phi_0^{\rm TE}}\left(\om^2+\Delta_{||}\right)\frac{e^{i\om|\x|}}{|\x|},
\nn\\ E_{\rm P}(\x) &=&e^{i\k\x}+\frac{1}{\phi_0^{\rm P}}\left(\om^2+\pa_z^2\right)\frac{e^{i\om|\x|}}{|\x|},
\label{2.3.30}\eea
which are the   expressions for the electric fields in the presence of one point dipole with polarizability according to \Ref{1.5}, provided proper definition of the $\phi_0's$ which will be given in the next subsection.

\subsection{Electrostatic approach}\label{s2.5}
In electrostatics, also if including retardation, one considers an applied electric field, in our case a plane wave,
\be \E_{\rm appl.}(\x)=\E_0\ e^{i\k\x}.
\label{2.4.1}\ee
It polarizes the dipole with dipole moment $\p$, located at the origin, whose electric field
$(\om^2+\nabla\circ\nabla)\, \frac{e^{i\om|\x|}}{4\pi|\x|}\p$,
%
together with $\E_{\rm appl.}(\x)$, adds up to the total electric field,
\be \E(\x)=\E_0\ e^{i\k\x}+(\om^2+\nabla\circ\nabla)\, \frac{e^{i\om|\x|}}{4\pi|\x|}\p\,.
\label{2.4.3}\ee
The dipole is assumed to be polarized by a local field, $\E_{\rm loc.}(\x)$, according to the second line in \Ref{1.2},
\be \p=\hat{\al}\E_{\rm loc.}(0),
\label{2.4.4}\ee
at the location of the dipole. Now, in case of a single dipole, the local field is assumed to coincide with the applied field at this position,
\be \E_{\rm loc.}(0)=\E_{\rm appl.}(0)
\label{2.4.5}\ee
and from \Ref{2.4.3} one comes to
\be \E(\x)=\E_{\rm appl.}(\x)+\left((\om^2+\nabla\circ\nabla)\,
    \frac{e^{i\om|\x|}}{4\pi|\x|}\right)\hat{\al}\E_{\rm appl.}(0).
\label{2.4.6}\ee
This is the electric field including the response of the dipole. It is to be mentioned that in this procedure no singularities appear and that the result is perfectly fine and in \Ref{2.4.6} the polarizability $\hat{\al}$ is the original one from eq. \Ref{1.2}.  For instance, from the second term one can derive the scattering amplitudes of Thomson scattering, Rayleigh scattering etc. The properties of the dipole enter through its polarizability and no ambiguity appears.

Now we compare this approach with the previously considered ones.
Inserting for $\hat{\al}$ from \Ref{1.5} and selecting with \Ref{1.7} the corresponding modes and components of the electric field, one comes to the conclusion that one needs to include the terms with $i\om^3$ in \Ref{2.3.26} into the renormalization, thus defining
\be\begin{array}{lrll}
    \frac{1}{\al_{||}^{\rm ren}}   &=&  \frac{1}{\al_{||}}-\frac{\om^2}{\sqrt{2\pi\ep}}+i\om^3,&\quad {\rm for~ TE,}\\[4pt]
     \frac{1}{\al_{||}^{\rm ren}}   &=&  \frac{1}{\al_{||}}+\frac{1}{6\sqrt{2\pi}\ep^{3/2}}-\frac{\om^2}{3\sqrt{2\pi\ep}}+\frac{i\om^3}{3},&\quad {\rm for~ TM,}\\[4pt]
      \frac{1}{\al_{3}^{\rm ren}}   &=&  \frac{1}{\al_{3}}+\frac{1}{12\sqrt{2\pi}\ep^{3/2}}-\frac{\om^2}{6\sqrt{2\pi\ep}}+\frac{2i\om^3}{3},&\quad {\rm for ~P,}
       \end{array}
\label{2.4.7}\ee
and one comes to
\be \phi_0^{\rm TE}=\frac{1}{\al_{||}^{\rm ren}},\quad
    \phi_0^{\rm TM}=\frac{1}{\al_{||}^{\rm ren}},\quad
    \phi_0^{\rm P}=\frac{1}{\al_{3}^{\rm ren}}
\label{2.4.8}\ee
in place of \Ref{2.3.27a} and for the $\al^{\rm ren.}$ we have to take the original ones appearing in \Ref{2.4.6} or, with \Ref{1.5} from \Ref{1.2}. As already mentioned, the renormalization is not unique.
In the approach of using an equation like \Ref{1.6a} with subsequent renormalization of the parameters $\hat{\al}$, like in the scalar case, these loose their meaning. The renormalized parameters $\al^{\rm ren.}$, \Ref{2.3.27}, can be related via equations \Ref{2.3.30} with the scattering amplitudes like in the scalar case. In the electrostatic approach, because of disregarding the self fields, one does not have an equation like \Ref{1.6a} but the system \Ref{1.2} with the prescription resulting in \Ref{2.4.6}. In this case no renormalization is needed for. It is, if comparing this approach with that of subsection \ref{s2.4}, that the renormalization \Ref{2.4.7} is suggested in order to obtain the same result.

It must be mentioned, that this electrostatic approach is a kind of workaround. It is a known fact, that equations \Ref{1.2}, considered together, are singular like eq. \Ref{1.6a}. The workaround is generally accepted and it gives in all known cases the correct result. For detail we refer to Chapt. 17 in \cite{Jackson75}.


\section{Multiple centers}\label{s3}
In this section we consider a plane lattice of delta functions at the locations
\be
    \x_\n=a\n,\quad\mbox{with}\quad    \n= (n_1,n_2,0),
\label{3.1}\ee
($n_i$ integer) forming a homogeneous lattice in the $(x,y)$-plane with spacing $a$. This lattice has translational invariance and, for instance, for the difference between 2 locations
\be \a_\n-\a_\m=\a_{\n-\m}
\label{3.2}\ee
holds.

We start with formulas with are valid for an arbitrary locations of the centers.
For the scalar field we have the equation
\be
    \left(-\om^2-\Delta+g\sum_\n \delta^3(\x-a_\n)\right)\Phi(\x)=0,
\label{3.3}\ee
which was already mentioned in the Introduction, eq. \Ref{1.1}. We use the approaches discussed in the preceding section for a single center. Eq. \Ref{3.3} is for the scalar field which we consider first. The extension to the electric field is given below.

In the method of selfadjoint extensions we have first to restrict the domain to functions regular the locations of all delta functions and then to add the singular modes. In this way, the solutions have   expansions around each delta function,
\be \Phi(\x)=c\left(\frac{1}{|\x-\a_\n|}+\al_{\rm SE}+O(|\x-\a_\n|)\right),
\label{3.4}\ee
with extension parameter $\al_{\rm SE}$. This is equivalent to use boundary conditions \Ref{2.1.2} at each $\a_\n$. We chose the same parameter $\al_{\rm SE}$ for all locations not to break the homogeneity of the lattice.

The solutions for $\x\ne\a_\n$, i.e., outside the centers, read
\be \Phi(\x)=e^{i\k\x}+\sum_\n f_\n\ \frac{e^{i\om|\x-\a_\n|}}{|\x-\a_\n|}.
\label{3.5}\ee
These may be interpreted a incoming plane wave and outgoing spherical waves from each center.  Expanding this solution near $\a_\m$,
\be \Phi(\x)=e^{i\k\a_\m}+f_\m\left(\frac{1}{|\x-\a_\m|}+i\om+\dots\right)
        +\sum_{\n\ne\m}f_\n\ \frac{e^{i\om|\a_\m-\a_\n|}}{|\a_\m-\a_\n|},
\label{3.6}\ee
and comparing with  \Ref{3.4}, we identify
\be \al_{\rm SE}=\frac{e^{i\k\a_\m}}{f_\m}+i\om+
        \sum_{\n\ne\m}\frac{f_\n}{f_\m}\ \frac{e^{i\om|\a_\m-\a_\n|}}{|\a_\m-\a_\n|},
\label{3.7}\ee
which is a system of equations for the $f_\n$,
\be (\al_{\rm SE}-i\om)f_\m
-\sum_{\n\ne\m} \frac{e^{i\om|\a_\m-\a_\n|}}{|\a_\m-\a_\n|}f_\n=e^{i\k\a_\m}.
\label{3.8}\ee
The inhomogeneous solution of this system, being inserted into \Ref{3.5}, delivers the solution describing the response to the applied plane wave. The homogeneous solution, if existing, describes intrinsic excitations or bound states, depending on the application considered.

Obviously, in the approach with the zero range potentials one comes to the same equations. A similar picture appears with the approach with regularization and renormalization which we consider now in more detail in terms of separable potentials. For several centers, the kernel \Ref{2.3.14} takes the form
\be K(\x,\y)=(-\om^2-\Delta)\delta(\x-\y)+g\sum_n\delta_1(\ep,\x-\a_\n)\delta_2(\ep,\a_\m-\y)
\label{3.9}\ee
and for the Green function one has
\be G(\x,\y)=G_0(\x-\y)-\sum_{\n,\m}\int dx'   dy'\ G_0(\x-\x')\delta_1(\ep,\x'-\a_\n)
        \Phi^{-1}_{\n,\m}\delta_2(\ep,\a_\m-\y')G_0(\y'-\y)
\label{3.10}\ee
with
\be \phi_{\n,\m}=\frac{1}{g}+\int d\x\,d\y\ \delta_2(\ep,\a_\m-\y)G_0(\y-\x)\delta_1(\ep,\x-\a_\m).
\label{3.11}\ee
Inserting \Ref{3.10} into \Ref{2.3.15} results in
\be \sum_\n\phi_{\n,\n'}\phi^{-1}_{\n',\m}=\delta_{\n,\m},
\label{3.12}\ee
which is analog to eq. \Ref{3.8} in the extension approach.

The generalization of the solution \Ref{2.3.18} corresponding to an incoming plane wave follows from eq. \Ref{3.10} in the same way as before from eq. \Ref{2.3.18},
\be \Phi(\x)=e^{i\k\x}-\sum_{\n,\m}\int dx'\,G_0(\x-\x')\delta_1(\ep,\x'-\a_\n)\phi^{-1}_{\n,\m}
        \int d\y\,\delta_2(\ep,\a_\m-\y)e^{i\k\y}.
\label{3.13}\ee
Taking $\ep\to0$ in the regularized delta functions in this formula, one comes to
\be \Phi(\x)=e^{i\k\x}-\sum_{\n,\m}\phi^{-1}_{\n,\m}e^{i\k\a_\m}\frac{e^{i\om|\x-\a_\n|}}{|\x-\a_\n|}.
\label{3.14}\ee
Comparison with eq. \Ref{3.5} gives
\be f_\n=\frac{-1}{4\pi}\sum_{\m}\Phi^{-1}_{\n,\m}e^{i\k\a_\m}
\label{3.15}\ee
establishing the relation to the extension method.

Starting from here we restrict the discussion to the plane lattice  given by eq. \Ref{3.1} and make use of the translational invariance \Ref{3.2}. In that case, solution \Ref{3.5} has a Bloch wave property,
\be \Phi(\x+\a_\n)=e^{i\k\a_\n}\Phi(\x).
\label{3.16}\ee
As a consequence,
\be f_\n=f_0e^{i\k\a_\n}
\label{3.17}\ee
holds and eq. \Ref{3.8} turns into
\be \left((\al_{\rm SE}-i\om)e^{i\k\a_\m}-
    \sum_{\n\ne\m}\frac{e^{i\om|\a_\m-\a_\n|+i\k\a_\n}}{|\a_\m-\a_\n|}\right)f_0=e^{i\k\a_\m},
\label{3.18}\ee
having simply the solution
\be f_0=\frac{1}{\al_{\rm SE}-i\om-J_1(\om,\k)}.
\label{3.19}\ee
Here we introduced the notation
\be J_s(\om,\k)=\sum_\n\,\frac{1}{|a_\n|^s}\,e^{i\om|\a_\n|+i\k\a_\n},
\label{3.20}\ee
which is the generic form of the sums appearing in the considered type of problems. For $\om=0$ and $\k=0$, $J_s(0,0)$ is an Epstein zeta function. In \Ref{3.19} the convergence of the sum comes from $\Im \om>0$.

With \Ref{3.19}, \Ref{3.17} and \Ref{3.15}, to solution \Ref{3.14} can be written as
\be \Phi(\x)=e^{i\k\x}+f_0 \,F_{\om,\k}(\x),
\label{3.21}\ee
where we introduced the notation
\be F_{\om,\k}(\x)=\sum_\n\frac{\exp(i\om|\x-\a_\n|+i \k \a_\n)}{|\x-\a_\n|},
\label{3.22}\ee
which is a weighted sum over the spherical waves outgoing from each center. In \Ref{3.21}, the coefficient $f_0$ can be rewritten using \Ref{2.1.5} and \Ref{3.19},
\be f_0=\frac{f}{1-f\,J_1(\om,\k)},
\label{3.23}\ee
where $f$ is the scattering amplitude for a single center. The denominator in \Ref{3.23} is clearly the result from multiple scattering within the lattice of delta functions.

Next we consider the corresponding formulas with the regularization in terms of separable potentials following from \Ref{3.10} and \Ref{3.11} making now use of the translational invariance \Ref{3.2}. For the regularized delta functions we use directly the heat kernel \Ref{2.3.21}. In this way we get
\be \Phi(\x)=e^{i\k\x}-\sum_{\n,\m}\int_0^\infty ds\, e^{-s\xi^2}
    K_{s+\ep}(\x-\a_\n) \Phi^{-1}_{\n,\m} e^{-\ep k^2+i\k\a_\m}
\label{3.24}\ee
for the solution \Ref{3.13}, where we turned again to imaginary frequencies like in \Ref{2.3.2}. Further we used
\be \int d\y\, K_\ep(\a_\m-\y)\, e^{i\k\y}=e^{-\ep k^2+i\k\a_\m}
\label{3.25}\ee
following from  \Ref{2.3.21}. From \Ref{3.11} we get
\be \phi_{\n,\m}=\frac{\delta_{\n,\m}}{g}+\int_0^\infty ds\,e^{-s\xi^2}K_{s+2\ep}(a_{\n-\m}),
\label{3.26}\ee
which is, in fact, a function of the difference, $\phi_{\n,\m}=\phi_{\n-\m,0}$. This allows to invert the matrix $\phi_{\n,\m}$ by Fourier transform. Defining
\be \tilde{\phi}(\k)=\sum_\n \phi_{\n,0}e^{i\k \a_\n}
\label{3.27}\ee
we get
\be \sum_\m \phi_{\n,\m}^{-1}e^{i\k\a_\m}=\frac{1}{\tilde{\phi}(\k)} \, e^{i\k\a_\n}
\label{3.28}\ee
and the solution \Ref{3.23} takes the form
\be \Phi(\x)=e^{i\k\x}-\frac{1}{4\pi\tilde{\phi}(\k)}F_{\ep,\om,\k}(\x)
\label{3.29}\ee
with
\be F_{\ep,\om,\k}(\x)=4\pi\sum_\n \int_0^\infty ds\,
        e^{-s\xi^2-\ep k^2+i\k\a_\n}    K_{s+\ep}(\x-\a_\n).
\label{3.30}\ee
This is the sum of the smeared (regularized) outgoing spherical waves from each center. For $\ep\to0$ it turns into \Ref{3.22} and the relation
\be f_0=\frac{-1}{4\pi\tilde{\phi}(\k)}
\label{3.30a}\ee
holds.

Next we have to consider the renormalization. It is hidden in eq. \Ref{3.26} in the diagonal contributions,
\be \phi_{\n,\n}=\frac{1}{g}+ \int_0^\infty ds\,e^{-s\xi^2}K_{s+2\ep}(0),
\label{3.31}\ee
which is just $\phi_0$ in eq. \Ref{2.3.24}. It is obvious that the divergent contributions come only from the diagonal elements due to the decrease of the heat kernel in \Ref{3.26} for $a_{\n-\m}\ne0$. In this way, all formulas of section II.3 related to the renormalization apply here too.

Especially for the scalar case we get then in \Ref{3.27} with \Ref{2.3.27}
\be \tilde{\phi}(\k)=\frac{1}{g_r}+\frac{i\om}{4\pi}+
    {\sum _\n}' \int_0^\infty ds\, e^{-s\xi^2}K_{s+2\ep}(\a_\n) \, e^{i\k\a_\n},
\label{3.32}\ee
where the prime at the sum means, as usual, to drop the term with $\n=0$. Here we can put $\ep=0$ and get
\be \tilde{\phi}(\k)=\frac{1}{g_r}+\frac{i\om}{4\pi}+\frac{J_1(\om,\k)}{4\pi}
\label{3.33}\ee
with $J_1(\om,\k)$ defined in \Ref{3.20}. Inserted into \Ref{3.29} and taken for $\ep=0$, this repeats just \Ref{3.21} with \Ref{3.19} and \Ref{2.3.10}.

The same can be done for the electric field using the modes corresponding to \Ref{1.8} and the renormalization according to eq. \Ref{2.4.7}. Basically, it amounts in inserting the operator $P$ defined in \Ref{1.8}, in front of the heat kernel in \Ref{3.32} and carrying out the derivatives. Using eq. \Ref{2.3.23}, we get, already for $\ep=0$,
\bea    \tilde{\phi}^{\rm TE}(\k) &=& \frac{1}{\al^{\rm ren}_{||}}+
             {\sum _\n}'\int_0^\infty ds\, \frac{e^{-s\xi^2}}{(4\pi s)^{3/2}}
                \ 4\pi \xi^2 \exp({-\frac{\a_\n^2}{4s}+i\k\a_\n}),
\nn\\      \tilde{\phi}^{\rm TM}(\k) &=& \frac{1}{\al^{\rm ren}_{||}}+
             {\sum _\n}'\int_0^\infty ds\, \frac{e^{-s\xi^2}}{(4\pi s)^{3/2}}
                \ 4\pi \left(\xi^2+\frac{1}{ s}-\frac{\a_\n^2}{4s^2}\right) \exp({-\frac{\a_\n^2}{4s}+i\k\a_\n}),
\nn\\      \tilde{\phi}^{\rm P}(\k) &=& \frac{1}{\al^{\rm ren}_{3}}+
             {\sum _\n}'\int_0^\infty ds\, \frac{e^{-s\xi^2}}{(4\pi s)^{3/2}}
                \ 4\pi \left(\xi^2+\frac{1}{2s} \right) \exp({-\frac{\a_\n^2}{4s}+i\k\a_\n}).
\label{3.34}\eea
Carrying out the integrations over $s$ and using the notations \Ref{3.20} we come to
\bea    \tilde{\phi}^{\rm TE}(\k) &=& \frac{1}{\al^{\rm ren}_{||}}-\om^2 J_1(\om,\k),
\nn\\      \tilde{\phi}^{\rm TM}(\k) &=& \frac{1}{\al^{\rm ren}_{||}}+i\om J_2(\om,\k)-J_3(\om,\k),
\nn\\      \tilde{\phi}^{\rm P}(\k) &=& \frac{1}{\al^{\rm ren}_{3}}-
                        \om^2J_1(\om,\k)-i\om J_2(\om,\k)+J_3(\om,\k),
\label{3.35}\eea
which are the formulas coming for the electric field in place of \Ref{3.33} for the scalar field.

Now the solutions for the electric field are given by eq. \Ref{3.29} with inserting $P$ from \Ref{1.8} in front of $F_{\ep,\om,\k}(\x)$, eq. \Ref{3.30}, which follows from \Ref{2.3.23} and the substitution of $\tilde{\phi}(\k)$ by that from \Ref{3.35} for the corresponding cases. Taking the limit $\ep\to0$, one comes to the formulas
\bea E_{\rm TE}(\x) &=& e^{i\k\x}+\frac{1}{ \tilde{\phi}^{\rm TE}(\k)}  \om^2
            F_{\om,\k}(\x)      \nn\\
        E_{\rm TM}(\x) &=& e^{i\k\x}+\frac{1}{ \tilde{\phi}^{\rm TM}(\k)}
        \left( \om^2+\Delta_{||}\right)  F_{\om,\k}(\x)      \nn\\
        E_{\rm P}(\x) &=& e^{i\k\x}+\frac{1}{ \tilde{\phi}^{\rm P}(\k)}
        \left( \om^2+\pa_z^2\right)  F_{\om,\k}(\x)
\label{3.36}\eea
where for the $\tilde{\phi}$ one needs to insert from \Ref{3.35}. We mention that the polarization vector usually appearing in front of the plane wave contribution is absorbed in \Ref{1.7} and does not appear here. In fact, $F_{\om,\k}(\x)$  with $P$ from \Ref{1.8} applied, are the well known formulas  for the electric field from a dipole, written here for the specific cases considered.

\section{Scattering on a \td lattice}\label{s4}
In this section we consider the solutions found in the preceding section for a plane lattice in a scattering setup in two cases, for outgoing spherical and plane waves. In both cases, the incoming wave is the same plane wave as before.

For the spherical setup one considers an outgoing spherical wave at $|\x|\to\infty$,
\be \Phi(\x)\raisebox{-4pt}{$\sim\atop |\x|\to\infty$}
        e^{i\k\x}+f_{sc}\frac{e^{i\om|\x|}}{|\x|},
\label{4.1}\ee
where $f_{sc}$ is the scattering amplitude. For a single center, the solution \Ref{2.1.3}, has already this form since  the second term in the solution is everywhere, except on the origin, a single outgoing spherical wave  and $f_{sc}=f$ holds.

For multiple centers \Ref{3.1} we have as usual for $|\x|\to\infty$,
\be |\x-\a_\n|=|\x|-a\hat{\x}\n+\dots\, ,
\label{4.2}\ee
where $\hat{\x}=\x/|\x|$ is the scattering direction. From \Ref{3.22} we get
\be F_{\om,\k}(\x)\raisebox{-4pt}{$\sim\atop |\x|\to\infty$}
\frac{e^{i\om|\x|}}{|\x|}\sum_\n  e^{-ia\om\hat{\x}\n+i a\k  \n},
\label{4.3}\ee
and from \Ref{3.21} and \Ref{4.1} the scattering amplitude
\be f_{sc}=f_0\sum_\n e^{-ia\om\hat{\x}\n+i a\k  \n}
\label{4.3a}\ee
follows. Doing Poisson resummation we get
\be f_{sc}=f_0  \left(\frac{2\pi}{a}\right)^2 \sum_\n
        \delta\left(\om\hat{\x}_{||}-\k_{||}+\frac{2\pi}{a}\n\right)
\label{4.4}\ee
with a \td delta function and the subscript '$||$' denotes the directions parallel to the plane. In this way, only scattering in the directions given by
\be \hat{\x}_{||}=\frac{1}{\om}\left(\k_{||}-\frac{2\pi}{a}\n\right)
\label{4.5}\ee
is allowed. These are just the v.Laue conditions. Within the zero range potential approach this was pointed out in \cite{karp81-19-58}.

Next we consider an outgoing plane wave. For this it is useful to rewrite $F_{\om,\k}(x)$, eq. \Ref{3.22}, using \Ref{2.3.17a},
\be F_{\om,\k}(x)=\sum_\n \int\frac{d \p}{2\pi^2}\ \frac{e^{i\p(\x-a\n)+i\k a\n}}{\p^2-\om^2-i0}\,.
\label{4.6}\ee
Doing again Poisson resummation and carrying out the integrations over $\p_{||}$,  one comes to
\be F_{\om,\k}(x)=\left(\frac{2\pi}{a}\right)^2 \sum_\n \int_{-\infty}^\infty \frac{d p_3}{2\pi^2}\
    \frac{\exp\left(i\left(\k_{||}+\frac{2\pi}{a}\n\right)\x_{||}+ip_3 z\right)}
            {\left(\k_{||}+\frac{2\pi}{a}\n\right)^2+p_3^2-\om^2-i0}.
\label{4.7}\ee
The integration over $p_3$ can be carried out too,
\be F_{\om,\k}(x)=\frac{4\pi}{a^2}\sum_\n \frac{i}{2\Gamma_n}\, e^{i\Gamma_n|z|+i\left(\k_{||}+\frac{2\pi}{a}\n\right)\x_{||}},
\label{4.8}\ee
where we defined
\be \Gamma_n=\sqrt{\om^2-\left(\k_{||}+\frac{2\pi}{a}\n\right)^2},
\label{4.9}\ee
inheriting a positive imaginary part, $\Im \Gamma_n>0$, from $\om$. Inserted into \Ref{3.21}, the solution becomes
\be \Phi(\x)=e^{i\k\x}+\frac{2\pi i f_0}{a^2} \sum_\n \frac{e^{i\Gamma_n|z|+i\left(\k_{||}+\frac{2\pi}{a}\n\right)\x_{||}}}{\Gamma_n}.
\label{4.10}\ee
Being interested in scattering solutions not decreasing for $|z|\to\infty$, one needs to have real $\Gamma_n$. This implies the restriction
\be |\n|\le\frac{a}{2\pi}(\om-k_{||}),
\label{4.11}\ee
thus restricting the summation in \Ref{4.10} to a finite sum.
A restriction to small $\n$ appears, for instance, for a small lattice spacing $a$, or, with $\om=\sqrt{k_{||}^2+k_3^2}$, for small $k_3$. However, scattering with $\n=0$ is always possible.
For the plane wave scattering setup on a homogeneous plane lattice it is useful to introduce Bloch waves in the parallel directions. Dividing the wave vector of the incoming plane wave,
\be k_{||}=\q+\frac{2\pi}{a}\m,
\label{4.12}\ee
where $\q$ is the quasiimpuls, $q_i \le \frac{\pi}{a}$ ($i=1,2$), we represent the solution in the form
\be \Phi(\x)=e^{i\q\x_{||}}\sum_\n\Phi_\n(z) \, e^{i\frac{2\pi}{a}\n\x_{||}}
\label{4.13}\ee
and get from \Ref{4.10}
\be \Phi_\n(z)=\delta_{\n,\m}e^{ik_3z}+\frac{2\pi if_0}{a^2}\frac{1}{\Gamma_{\n-\m}}
                \,e^{i\Gamma_{\n-\m}|z|}.
\label{4.14}\ee
Here $\m$ denotes the zone of the incoming wave. The response has scattered waves with, in general, all numbers $\n$. From considering this solution  for $|z|\to\infty$, it makes sense to define the reflection coefficients
\be r_{\n}=\frac{2\pi i f_0}{a^2\Gamma_{\n}},
\label{4.15}\ee
or, with \Ref{3.30a},
\be r_{\n}=\frac{-i}{2a^2\Gamma_{\n}\tilde{\phi}(k)},
\label{4.15a}\ee
and $\Gamma_{\n}$ \Ref{4.9} is the wave number of the scattered wave moving in $z$-direction \Ref{4.10}. Since we have to account here only for functions not decreasing at infinity,
the restriction on $\n$ to such values for which $\Gamma_{\n}$ is real, holds as before.

Finally we consider the electric field. The structure of the formulas for the scattering into as plane outgoing wave remains the same except for the modifications coming from the $P$'s in \Ref{1.8} and for $f_0$ for which we have to use $f_0\to -1/(4\pi\tilde{\phi}(\k)$ in \Ref{3.29} with $\tilde{\phi}(\k)$ from \Ref{3.35}. Carrying out the derivatives in $P$ under the sign of the summation, we get in place of \Ref{4.10}
\bea E_{\rm TE}(\x) &=& e^{i\k\x}-\frac{2\pi i}{ a^2 \tilde{\phi}^{\rm TE}(\k)}
        \sum_\n  \om^2 \ \frac{e^{i\Gamma_\n|z|+i(\k_{||}+\frac{2\pi}{a}\n)\x_{||}}}{\Gamma_\n}  , \nn\\
 E_{\rm TM}(\x) &=& e^{i\k\x}-\frac{2\pi i}{ a^2 \tilde{\phi}^{\rm TM}(\k)}
        \sum_\n  \left(\om^2- (\k_{||}+\frac{2\pi}{a}\n)^2\right)\ \frac{e^{i\Gamma_\n|z|+i(\k_{||}+\frac{2\pi}{a}\n)\x_{||}}}{\Gamma_\n} ,  \nn\\
 E_{\rm P}(\x) &=& e^{i\k\x}-\frac{2\pi i}{ a^2 \tilde{\phi}^{\rm P}(\k)}
        \sum_\n  \left(\om^2- \Gamma_\n^2+i\Gamma_\n \delta(z)\right)\ \frac{e^{i\Gamma_\n|z|+i(\k_{||}+\frac{2\pi}{a}\n)\x_{||}}}{\Gamma_\n}.
\label{4.16}\eea
The delta function in the last line is the one known for the normal component of the electric field on a double layer. It does not enter, of course, the asymptotics for $|z|\to\pm\infty$, which we are interested in. Eq. \Ref{4.16} can simplified a bit using \Ref{4.9}.

Introducing in parallel to \Ref{4.13} the corresponding Bloch waves, one comes to the reflection coefficients  given by
\bea    r^{\rm TE}_{\n}  &=&  \frac{-2\pi i}{a^2 \tilde{\phi}^{\rm TE}(\k) \Gamma_\n}\om^2,
\nn\\   r^{\rm TM}_{\n}  &=&  \frac{-2\pi i}{a^2 \tilde{\phi}^{\rm TM}(\k)} \Gamma_\n ,
\nn\\   r^{\rm P}_{\n}  &=&  \frac{-2\pi i}{a^2 \tilde{\phi}^{\rm P}(\k) \Gamma_\n}
                           (\k_{||}+\frac{2\pi}{a}\n)^2,
\label{4.17}\eea
which come in place of \Ref{4.15}.

\section{The transition to a continuous sheet}\label{s5}
Formally, for vanishing lattice spacing, $a\to 0$, the scattering centers become dense and form a continuous sheet. However, this limiting process is quite singular and unexpected results may appear.
First of all we mention that in the solution \Ref{4.10}, for $a\to 0$ only $n=0$ delivers a nonvanishing solution. From \Ref{4.9}, using $\om^2=\k_{||}^2+k_3^2$, we have then $\Gamma_0=k_3$ and this solution is
\be \Phi(\x)=e^{i\k_{||}\x_{||}}\left(e^{ik_3z}+r\, e^{i k_3|z|}\right)
\label{5.1}\ee
with
\be r=\frac{2\pi if_0}{a^2 k_3},
\label{5.2}\ee
or $r=r_0$ from \Ref{4.15}. Here, the factor $\frac{1}{a^2}=\rho$ has to be interpreted as density, which must be kept finite.

Using eq. \Ref{3.23} for $f_0$ and,
\be J_1(\om,\k) =\frac{2\pi i}{a^2 k_3}\quad\mbox{for}\quad a\to0,
\label{5.3}\ee
which follows from \Ref{3.20} by substituting, in leading order, the summation by integration, we get
\be r=\frac{-1}{1-\frac{k_3 a^2}{2\pi i   f}}.
\label{5.4}\ee
Further we insert  from \Ref{2.3.10},
\be \frac{f}{a^2}=\frac{-1}{\frac{4\pi a^2}{g_r}+i a^2 \om^2},
\label{5.5}\ee
where the second term in the denominator vanishes for $a\to0$ since in the first term $a^2/g_r$ must be kept finite. Using that in \Ref{5.4}, we get
\be r=\frac{-1}{1-\frac{2ik_3}{\rho g_r}},
\label{5.6}\ee
which is the reflection coefficient $r_0$, \Ref{4.15}, resulting from the solution derived in the preceding section for $a\to0$.

The transition $a\to0$ can also be done in the equation \Ref{3.3}. Thereby we ignore the problems that the delta function in this equation  is singular and, thus, that this transition has a rather formal character.
The term with the delta functions can be written as
\be g\sum_n\delta(\a-\a_\n)=g\sum_\n\delta(\x_{||}-a\n)\delta(z).
\label{5.7}\ee
For $a\to0$, the summation index $\n$ becomes continuous and the sum turns into an integral, which can be carried out,
\be g\sum_\n\delta(\x_{||}-a\n)\raisebox{-4pt}{$ \sim\atop a\to0$}
        g\int d\n \ \delta(\x_{||}-a \n)=\frac{g}{a^2}.
\label{5.8}\ee
As said above, the $\rho=1/a^2$ is interpreted as finite density. With \Ref{5.8}, the equation \Ref{3.3} turns into
\be (-\om^2-\Delta+g\rho \delta(z))\Phi(\x)=0.
\label{5.9}\ee
It has now a one-dimensional delta function potential and is well defined. Taking Fourier transform,
\be \tilde{\Phi}_{\k_{||}}(z)=\int d\x_{||} e^{-i\k_{||} \x_{||}} \Phi(\x),
\label{5.10}\ee
the scattering solution is
\be \tilde{\Phi}_{\k_{||}}(z)=e^{ik_3 z}+r\, e^{ik_3|z|}
\label{5.11}\ee
with
\be r=\frac{-1}{1-\frac{2ik_3}{g\rho}}.
\label{5.12}\ee
These formulas are well known, we use those from the Appendix in \cite{bord14-89-125015}.
Comparison with \Ref{5.6} shows the reflection coefficients are the same provided the relation $g=g_r$ holds. This way, taking the limit $a\to0$ first in the equation, or first solving the equation and taking the limit afterwards delivers the same reflection coefficient and it provides another relation for the renormalized coupling constant, and by means of \Ref{2.1.5}, it provides a relation to the parameter of the self adjoint extension.

For the electric field we start from eq. \Ref{4.16} and consider first the TE mode. Again, for $a\to0$, only the term with $\n=0$ survives at $|z|\to\infty$ in the sum and the reflection coefficients \Ref{4.17} become $r_0\to r$ like in \Ref{5.2}. For the TE mode we get from the first line in \Ref{4.17}
\be r^{\rm TE}=\frac{-2\pi i \om^2}{a^2\tilde{\phi}_0(\k)k_3}
\label{5.13}\ee
with
\be \tilde{\phi}_0(\k)=\frac{1}{\al^{\rm ren.}_{||}}-\frac{2\pi i \om^2}{a^2 k_3},
\label{5.14}\ee
which can be rewritten in the form
\be r^{\rm TE}=\frac{1}{1-\frac{a^2k_3}{2\pi i \om^2 \al^{\rm ren.}_{||}}}.
\label{5.15}\ee
With this reflection coefficient, the scattering solution is the same as \Ref{5.1} with $E^{\rm TE}(\x)$ in place of $\Phi(\x)$. Equally well it can be obtained from \Ref{4.16} for $a\to0$.

Like in the scalar case we consider for comparison the transition $a\to0$ in the equation,
\be \left(-\om^2-\Delta-4\pi \al_{||}\om^2\sum_\n\delta(\x-\a_\n)\right)E_{\rm TE}(\x)=0,
\label{5.15a}\ee
following from \Ref{1.6b} with \Ref{1.8}. Doing the same operations as in eqs. \Ref{5.7} and \Ref{5.8}, we get
\be \left(-\om^2+\k_{||}^2-\pa_z^2-4\pi \al_{||}\om^2\delta(z)\right)\tilde{E}_{k_{||}}^{\rm TE}(z)=0,
\label{5.16}\ee
where we also introduced the Fourier transform like in \Ref{5.10}. We get the same equation as \Ref{5.9} with the change $g\to-4\pi\a_{||}\om^2$. This same substitution, as seen, connects \Ref{5.12} with \Ref{5.15}. Thus for the TE mode we have, like in the scalar case, a well defined transition to the continuous sheet. These formulas coincide with the hydrodynamic model and with \cite{bord14-89-125015}, which is an expected result.

Now we come to the TM modes for both, parallel and perpendicular polarizabilities. The solutions for a lattice with finite $a$ involve $\tilde{\phi}^{\rm TM}(k)$ and  $\tilde{\phi}^{\rm P}(k)$ which are given by eqs. \Ref{3.35}. For $a\to0$, the lattice sum $J_3(\om,k)$, \Ref{3.20}, converges and behaves as
\be J_3(\om,\k)\raisebox{-4pt}{$\sim\atop a\to0~$}\frac{1}{a^3}.
\label{5.17a}\ee
Consequently,
\be \tilde{\phi}^{\rm TM}(k)\raisebox{-4pt}{$\sim\atop a\to0~$}\frac{1}{a^3},
\qquad \tilde{\phi}^{\rm P}(k)\raisebox{-4pt}{$\sim\atop a\to0~$}\frac{1}{a^3}
\label{5.17}\ee
hold. This behavior is by one power of $1/a$ too singular. A factor $1/a^2$ can be absorbed into the density as done before, but  any additional factor cannot.
As a consequence, we do not get any sensible result for $a\to0$.
This singularity was pointed out in \cite{bart13-15-063028} for the perpendicular polarizability. A somehow unexpected result is that it appears also for one of the polarizations (TM) in case of parallel polarizability (which was not considered in \cite{bart13-15-063028}).

Let us consider the transition $a\to0$ also on the level of the equation \Ref{1.6b}. For the TM mode, the formal transition using eq. \Ref{5.8} cannot be done directly because of the derivatives in $\Delta_{||}$. We attempt to handle this by Fourier transform,
\be E_{\rm TM}(\x)=\int \frac{d\k_{||}}{(2\pi)^2}\, e^{i\k_{||}\x_{||}}
        \tilde{E}^{\rm TM}_{k_{||}}(z)
\label{5.18}\ee
and get
\be \left(-\om^2+k_{||}^2+\pa_z^2\right)\tilde{E}^{\rm TM}_{k_{||}}(z)
        -4\pi\al_{||}\sum_n\left(\om^2-(k_{||}+\frac{2\pi}{a}\n)^2\right)
                        \tilde{E}^{\rm TM}_{k_{||}+\frac{2\pi}{a}\n}(0)\ \delta(z)=0,
\label{5.19}\ee
where we used
\be \int d\x_{||} e^{-i\k_{||}\x_{||}}(\om^2+\Delta_{||})\sum_\n \delta(\x_{||}-a\n)E^{\rm TM}(0)
=\left(\frac{2\pi}{a}\right)^2\sum_\n \left(\om^2-(k_{||}+\frac{2\pi}{a}\n)^2\right)
\tilde{E}^{\rm TM}_{k_{||}+\frac{2\pi}{a}\n}(z)
.
\label{5.20}\ee
As expected, the derivatives turned into the corresponding momenta.

If now doing the transition $a\to0$, in case the solution $ \tilde{E}^{\rm TM}_{k_{||}+\frac{2\pi}{a}\n}(0)$ decreases for $\n\ne0$, only the term with $\n=0$ survives. This is just what happened for the scalar case and for the TE mode, where it resulted in the well known result  \Ref{5.9}.

The decrease of $ \tilde{E}^{\rm TM}_{k_{||}+\frac{2\pi}{a}\n}(0)$ can be guessed by solving the equation as in Sect. II.C. In doing so for the TM case, from the factor $\left(\om^2-(k_{||}+\frac{2\pi}{a})^2\right)$ one observes that this decrease is not present for the TM mode. A similar observation can be made if doing Fourier transform \Ref{5.18} with the solution \Ref{4.16}. It is again the factor $\left(\om^2-(k_{||}+\frac{2\pi}{a}\n)^2\right)$, which the TM mode has more, which prevents the decrease. As a consequence, a transition $a\to0$ in eq. \Ref{5.19} does not restrict the summation to $\n=0$.

It is to be mentioned that the restriction to $\n=0$, if doing it nevertheless, results in the equation
\be \left(-\om^2+\k_{||}^2-\pa_z^2
        -4\pi \al_{||}(\om^2-\k_{||}^2)\delta(z)\right)E^{\rm TM}(\x)=0
\label{5.21}\ee
for the TM mode. The scattering solution, written in parallel  to \Ref{5.1}, has a reflection coefficient
\be r_{\rm TM}=\frac{-1}{1-\frac{a^2}{2\pi i \al_{||} k_3}},
\label{5.22}\ee
where we used $\om^2=\k_{||}^2+k_3^2$. This equation and this reflection coefficient are there same as in the hydrodynamic model for a continuous sheet.

A similar picture appears for the TM mode in the case of perpendicular polarizability. The equation following formally from \Ref{1.6b} is
\be \left(-\om^2-\Delta-4\pi \al_{3}(\om^2+\pa_z^2)\sum_\n\delta(\x-\a_\n)\right)E_{\rm P}(\x)=0.
\label{5.23}\ee
After doing Fourier transform \Ref{5.18}, it becomes
\be \left(-\om^2+k_{||}^2-\pa_z^2\right)\tilde{E}^{\rm P}_{k_{||}}(z)
        -4\pi\al_{3}\sum_n\left(\om^2-\pa_z^2\right)
                        \tilde{E}^{\rm P}_{k_{||}+\frac{2\pi}{a}\n}(0)\delta(z)=0,
\label{5.24}\ee
Here the problem is in the second derivative of the delta function, $\delta''(z)$, which appears in addition to \Ref{5.21}. How this problem could be handled was shown in \cite{bord14-89-125015} by going beyond the dipole approximation. This is, however, not topic of the present paper.
In addition to $\delta''(z)$, eq. \Ref{5.24} has for $a\to0$ also the problem that the restriction to $\n=0$ does not appear either. Here the  missing decrease of
$\tilde{E}^{\rm P}_{\k_{||}+\frac{2\pi}{a}\n}(0)$ for $\k_{||}\to\infty$ cannot be seen directly from the equation as before, however from the solution \Ref{4.16}. It has with $\Gamma_\n$ a nondecreasing factor.

In this way, in both TM cases, the derivatives, resulting from those in the first line in the right sides of eq. \Ref{1.2}, prevent a transition to the continuous sheet.

\section{Transition to continuous sheet keeping a regularization}\label{s6}
We have seen that after renormalization, the regularization can be removed, $\ep\to0$, delivering meaningful results. These agree with the electrostatic approach where the self fields are excluded. Further we have seen that a part of these results becomes singular when the lattice spacing becomes small, $a\to0$. In this section we ask the question what happens if we first take $a\to0$, keeping $\ep>0$. The result should be a sheet of finite thickness given by $\ep$. Further we consider what happens if we take, after $a\to0$, also $\ep\to0$.

We start with the scalar case. The solution is given by \Ref{3.29}. For $F_{\ep,\om,\k}(\x)$ we use \Ref{3.30} and \Ref{2.3.21}. For $a\to0$, the summation over $\n$ turns into an \td integration according to
\be \sum_\n g(a\n) \to \frac{1}{a^2}\int d\n \,g(\n),
\label{6.1}\ee
for some function $g(\n)$, and we get
\be \Phi(\x)=e^{i\k\x}+\frac{1}{2ia^2k_3\tilde{\phi}(\k)}\, h_\ep(z)\, e^{i\k_{||}\x_{||}}
\label{6.2}\ee
with
\be h_\ep(z)=\frac{-ik_3}{\sqrt{\pi}}\int_0^\infty \frac{ds}{\sqrt{s+\ep}}
   \ \exp\left(-\frac{z^2}{4(s+\ep)}-s(\xi^2+\k_{||}^2)-2\ep\k_{||}^2\right).
\label{6.3}\ee
We accounted for $\xi=-i\om$, which gives $\sqrt{\xi^2+\k_{||}^2}=-ik_3$. The function $h_\ep(z)$ describes  the $z$-dependence on the solution to the regularized equation. It is decreasing,
\be h_\ep(z)\raisebox{-4pt}{$\sim\atop |z|\to\infty$}e^{-z^2/4\ep},
\label{6.4}\ee
such that the separable regularization gives a sheet of finite thickness without scattering.
It is only for $\ep=0$ that it turns into the scattered plane wave,
\be h_0(z)=e^{ik_3|z|}.
\label{6.5}\ee
This property gives rise to the definition
\be r_\ep=\frac{1}{2ia^2k_3\tilde{\phi}(\k)}
\label{6.6}\ee
of a reflection coefficient such that the solution can be written in the form
\be \Phi(\x)=e^{i\k\x}+r_\ep\, h_\ep(z)\, e^{i\k_{||}\x_{||}}.
\label{6.6a}\ee
In this way, the wave function, if first taking $a\to0$ and subsequently $\ep\to0$, turns into the same as when doing the limits the other way round, eq. \Ref{5.1}, with $r_\ep$ in place of $r$.

Actually, the same happens with $\tilde{\phi}(\k)$. entering \Ref{6.6}. From \Ref{3.32} we get with \Ref{6.1}
\be \tilde{\phi}(\k)_{|\a\to0}=\frac{1}{g^r}+\frac{1}{a^2\sqrt{4\pi}}\int_0^\infty\frac{ds}{\sqrt{s+2\ep}}
    \, e^{-s(\xi^2+\k_{||}^2)-2\ep\k_{||}^2}.
\label{6.7}\ee
The term $i\om/4\pi$ present in \Ref{3.32} is subleading for $a\to0$ and was dropped. If now in \Ref{6.7} taking $\ep\to0$, we get
\be \left(\tilde{\phi}(\k)_{|a=0}\right)_{|\ep=0}=\frac{1}{g^r}+\frac{i}{2a^2k_3},
\label{6.8}\ee
which gives in \Ref{6.6} just the same reflection coefficient as given by \Ref{5.6}. Thus, for $\tilde{\phi}(\k)$, the limits commute too.

Now we consider the TE case for the electric field. Going through the same formulas which resulted in \Ref{3.36}, but keeping $\ep$, we get
\be E^{\rm TE}(\x)=e^{i\k\x}+r_\ep^{\rm TE}\, h_\ep(z)\, e^{i\k_{||}\x_{||}}
\label{6.9}\ee
with
\be r_\ep^{\rm TE}=\frac{2\pi i \om^2}{a^2k_3\tilde{\phi}^{\rm TE}(\k)}
\label{6.10}\ee
and
\be \tilde{\phi}^{\rm TE}(\k)=\frac{1}{\al^{\rm ren.}_{||}}-\frac{\sqrt{4\pi}\om^2}{a^2}
    \int_0^\infty\frac{ds }{\sqrt{s+2\ep}}
    \, e^{-s(\xi^2+\k_{||}^2)-2\ep\k_{||}^2}.
\label{6.11}\ee
It can be seen that this integral is finite for $\ep\to0$ and results in the same reflection coefficient \Ref{5.15} as before. Thus, also for the TE mode, the limits do commute.

For the TM mode we get the same way as before from \Ref{3.36}
\be E^{\rm TM}(\x)=e^{i\k\x}+r_\ep^{\rm TM}\, h_\ep(z)\, e^{i\k_{||}\x_{||}}
\label{6.12}\ee
with
\be r_\ep^{\rm TM}=\frac{2\pi i k_3}{a^2 \tilde{\phi}^{\rm TM}(\k)}.
\label{6.13}\ee
This formula appears in such simple way since the derivatives in $\Delta_{||}$ act only on $e^{i\k_{||}\x_{||}}$. A slightly more complicated picture we observe  in $\tilde{\phi}^{\rm TM}(\k)$. Here we have from \Ref{3.34}, restoring $\ep$, for finite $a$,
\be \tilde{\phi}^{\rm TM}(\k)=
    \frac{1}{\al^{\rm ren.}_{||}}-\frac{1}{\sqrt{4\pi}}  {\sum_\n}'
    \int_0^\infty\frac{ds\, e^{-s\xi^2}}{(s+2\ep)^{3/2}}
    \left(\om^2-\frac{1}{s+2\ep}+\frac{\a_n^2}{4(s+2\ep)}\right)
    \, e^{-\frac{\a_\n^2}{4(s+2\ep)}+i\k_{||}\a_\n}.
\label{6.14}\ee
As long as $\ep>0$, we can use \Ref{6.1}. For $\ep=0$ we could not do that because of the singularity at $s=0$. This is, basically, the moment where the limits do not commute.

Using \Ref{6.1} in \Ref{6.14}, we get
\be \tilde{\phi}^{\rm TM}(\k)_{|\a\to0}=
     \frac{1}{\al^{\rm ren.}_{||}}-\frac{ \sqrt{4\pi}k_3^2}{a^2}
     \int_0^\infty\frac{ds}{\sqrt{s+2\ep}}
    \, e^{-s(\xi^2+\k_{||}^2)-2\ep\k_{||}^2},
\label{6.15}\ee
which has also a finite limit for $\ep\to0$,
\be \left(\tilde{\phi}^{\rm TM}(\k)_{|a=0}\right)_{|\ep=0}=
     \frac{1}{\al^{\rm ren.}_{||}}-\frac{2\pi i k_3}{a^2}
\label{6.16}\ee
(we remind that the factor $a^2$ is the inverse density and should be considered as finite). Comparison with \Ref{5.17} shows now the difference  resulting from the orders of the limits.

The reflection coefficient  following from \Ref{5.16} is just the same as \Ref{5.22} and, thus, in the hydrodynamic model, provided one makes in \Ref{6.16} the choice $\al_{||}^{\rm ren.}=\al_{||}$.

Finally, we consider the TM mode for the perpendicular polarizability. We get from \Ref{3.36}
\be E^{\rm P}(\x)=e^{i\k\x}+\frac{2\pi i}{a^2 k_3 \tilde{\phi}^{\rm P}(\k)}
    \left(\om^2+\pa_z^2\right)\, h_\ep(z)\, e^{i\k_{||}\x_{||}}.
\label{6.17}\ee
Again, for $\ep>0$ we have a decreasing function of $z$. For $\ep\to0$ it turns into
\be \lim_{\ep\to0} \left(\om^2+\pa_z^2\right)\, h_\ep(z) =
\left(\k_{||}^2-2\delta(z)\right)\, e^{ik_3|z|}.
\label{6.18}\ee
This way, the delta function in the normal component of the electric field returns. However, it does not influence scattering and we can define a scattering coefficient,
\be r_\ep^{\rm P}=\frac{2\pi i k_{||}^2}{a^2 k_3\tilde{\phi}^{\rm P}(\k)},
\label{6.19}\ee
like in the previous cases. The function $\tilde{\phi}^{\rm P}(\k)$ entering here follows from \Ref{3.34} using \Ref{6.1}
\be \tilde{\phi}^{\rm P}(\k)_{|\a\to0}=
     \frac{1}{\al^{\rm ren.}_{3}}-\frac{ \sqrt{4\pi}}{a^2}
     \int_0^\infty\frac{ds}{\sqrt{s+2\ep}}
     \left(\om^2-\frac{1}{2(s+2\ep)}\right)
    \, e^{-s(\xi^2+\k_{||}^2)-2\ep\k_{||}^2}.
\label{6.20}\ee
However, this function, in opposite to \Ref{6.15}, does not have a finite limit for $\ep\to0$,
\be \tilde{\phi}^{\rm P}(\k)= \frac{1}{\al^{\rm ren.}_{3}}
    -\frac{2\pi i \k_{||}^2}{a^2 k_3}  +\frac{\sqrt{2\pi}}{a^2 \sqrt{\ep}}+O(\sqrt{\ep}).
\label{6.21}\ee
In this way, for perpendicular polarizability, taking the limit $a\to0$ first, does not help.

It is interesting to remark that, when removing the singular term in \Ref{6.21} by hand, or by including it into $\al^{\rm ren.}_{3}$, insertion of \Ref{6.21} into \Ref{6.19} gives
\be r^{\rm P}=\frac{-1}{1-\frac{a^2 k_3}{2\pi i k_{||}^2 \al_3^{\rm ren.}}},
\label{6.22}\ee
which is just the reflection coefficient obtained in \cite{bord14-89-125015}, eq. (48) (up to a redefinition of $\al$ by $4\pi$). This is, given the singularities, a rather formal coincidence.

\section{Conclusions}\label{s7}
In the preceding sections we considered all known approaches to a single \dd delta function potential and a two dimensional lattice of such potentials. There are basically two approaches, the self adjoint extension method and regularization/renormalization. Although different, both approaches are related and deliver the same results. We reviewed these relations in detail as well as their relation to the 'zero range potential' method in quantum mechanics. As a nice technical tool we reviewed also the regularization in terms of separable potentials.

The initial motivation for this paper comes from \td polarizable sheets in electrodynamics. There the situation is a bit different. Usually one starts from the system \Ref{1.2} of coupled equations for the electric field in the presence of polarizable point dipoles and for the dipole moment induced by the electric field in dipole approximation. Inserting the dipole moment into the equation for the electric field, an equation with \dd delta function potential results. In the generally accepted electrostatic approach, as discussed in Sect. \ref{s2.5}, the last step is accompanied by dropping the self fields of the dipoles. It is shown, this is equivalent to a specific choice of renormalization. In fact, by this procedure, the equations alone, i.e., without dropping the self fields, do not form a closed set. As discussed in \cite{Jackson75}, Chapt. 17, this procedure gives the correct result in almost all cases. To our opinion, it is an open question how this procedure can be justified as a limiting case from full quantum electrodynamics.

Using any of the above approaches, the problem of the interaction of the electromagnetic field with a lattice of dipoles can be formulated correctly. For example, within a scattering setup, the wave functions and the reflection coefficients can be written down (Sect. \ref{s4}) in the form of known lattice sums, eq. \Ref{3.20}. Formulas like \Ref{3.35} and \Ref{4.16}, \Ref{4.17} allow to calculate any quantity in question, plasmons or the Casimir force between two parallel lattices for instance. Also it should be possible to calculate the quantum mechanical bound levels on a lattice generalizing \cite{bere85-33-2122}, where this was done for a \od lattice of zero binding energy.

In the limit of vanishing lattice spacing, $a\to0$, one comes formally to a continuous polarizable sheet. However, when using a regularization, there is a second limit. For example, in the regularized equation \Ref{2.3.12} it is $\ep\to0$ and a finite $\ep$ corresponds to a  smearing out of the delta functions. As it is easy to imagine, these limits do not commute in general. It is only for a scalar field and for the electromagnetic TE polarization in case of a sheet with parallel polarizability, that these limits do commute. For TE polarization,  a result comes out  which is known from the hydrodynamic model for a continuous sheet. For the TM polarization in both cases of polarizability, parallel and perpendicular to the sheet, these limits do not commute.

In the case of first doing the limit $\ep\to0$ of removing the regularization first, for the TM polarization the limit $a\to0$ is singular as seen in Eqs. \Ref{5.17a} and \Ref{5.17} for the refl;ection coefficients. For the perpendicular polarizability this was observed in \cite{bart13-15-063028}, for the parallel polarizability this is new and it was not expected. One can discuss that the reason for this singularity does not depend on the direction of the polarizability but results from the derivatives in the couplings in Eq. \Ref{1.8}.

It must be mentioned that for parallel polarizability, the equations turn formally into that of the hydrodynamic model. Actually, this is the reason to expect for the TM polarizability a finite result. We showed at which place it is impossible to do $a\to0$ in the equations and where  the difference between TE and TM polarizations shows up.

In the case of first doing the limit $a\to0$ of vanishing lattice spacing first, for the TM polarization in case of parallel polarizability, one comes to a well defined reflection coefficient and the subsequent limit $\ep\to0$ gives the result expected from the hydrodynamic model. This result is a bit counterintuitive since here the smeared delta functions largely overlap forming a  sheet of finite thickness. This sheet is quite specific due to the separable regularization. For a sheet of finite thickness with homogeneous permittivity it is known (see, for instance Sect. V in \cite{bord14-89-125015}) that the limit of thickness to zero must be accompanied by an increase of the polarizability (or permittivity) in order to get a nonzero result. In the separable regularization, considered here, this is not the case.

A similar picture appears for the perpendicular polarizability (here we have only the TM polarization). However, in addition here a singularity appears which results from the delta function the normal component of the electric field has on a double layer. By removing this singularity by hand, the same result appears which was obtained in \cite{bord14-89-125015} by stepping back from the dipole approximation.

In summary, for the electromagnetic field, the results depend on the order of taking the different limits involved. We consider this as a hint to go beyond the dipole approximation.

\bibliography{C:/Users/bordag/WORK/Literatur/bib/papers,C:/Users/bordag/WORK/Literatur/Bordag,C:/Users/bordag/WORK/Literatur/bib/libri}

\end{document}